\documentclass[aps,prx,twocolumn,floatfix,superscriptaddress,longbibliography]{revtex4-1}

	\usepackage{amsmath}
	\usepackage{amssymb}
	\usepackage{amsthm} 
	\usepackage{nicefrac}
	\usepackage{braket}
	\usepackage{enumerate}
	\usepackage[pdftex]{graphicx}
	\usepackage{txfonts}
	\usepackage{dcolumn} 
	\usepackage{color}
	\usepackage{bm}
	\usepackage[bottom]{footmisc}
	\usepackage[version=3]{mhchem}
	\usepackage[colorlinks,citecolor=blue,linkcolor=blue,urlcolor=blue]{hyperref}
	\newcommand{\avg}[1]{\left< #1 \right>} 
	
\begin{document}
	\title{Topological Hall effect in the Shastry-Sutherland lattice}
	\author{Munir Shahzad}
	\affiliation{Department of Physics and Physical Oceanography, Memorial University of Newfoundland, St.  John’s, Newfoundland \& Labrador A1B 3X7, Canada}
	\author{Nyayabanta Swain}
	\author{Pinaki Sengupta}
	\affiliation{School of Physical and Mathematical Sciences, Nanyang Technological University, 21 Nanyang Link, Singapore 637371}
	\date{\today}

\begin{abstract}
We study the classical Heisenberg model on the geometrically frustrated Shastry-Sutherland (SS) lattice 
with additional Dzyaloshinskii-Moriya (DM) interaction in the presence of an external magnetic field. 
We show that several noncollinear and noncoplanar magnetic phases,
such as the flux, all-in/all-out, 3in-1out/3out-1in, and canted-flux phases are stabilized over wide ranges of parameters 
in the presence of the DM interaction. We discuss the role of DM interaction 
in stabilizing these complex magnetic phases. 
When coupled to these noncoplanar magnetic phases, itinerant electrons experience  a finite Berry phase, 
which manifests in the form of topological Hall effect, whereby a non-zero transverse 
conductivity is observed even in the absence of a magnetic field.
We study this anomalous magneto-transport by calculating the electron band structure and transverse conductivity 
for a wide range of parameter values, and demonstrate the existence of topological Hall effect in the SS lattice. We explore the role of the strength of itinerant electron-local moment coupling on electron transport and show that the topological Hall features evolve significantly from strong to intermediate values of the coupling strength, and are accompanied by the appearance of a finite spin Hall conductivity.
\end{abstract}

\maketitle
	
\section{Introduction}\label{sec:intro}

The interplay of charge and spin degrees of freedom manifests in novel phases in strongly correlated electron systems~\cite{dagotto_colossal_2001,buhler_magnetic_2000,ozawa_vortex_2016,reja_coupled_2015}. 
One of the basic models that describes this interplay is 
the Kondo lattice model or the double exchange (DE) model, 
in which localized magnetic moments are coupled to itinerant electrons~\cite{PhysRev.82.403,PhysRev.100.675,JPSJ.64.2734,PhysRevLett.80.845,JPSJ.33.21}.
In these systems the conduction electrons and localized spins
affect each other in a self-consistent way. The mobile electrons mediate effective interactions between the localized spins, 
and dictate the magnetic behavior. On the other hand, the scattering of the mobile electrons from these localized moments 
decides the resulting electronic and transport properties of the system.
This interplay becomes more interesting, when the localized moments are arranged on a 
geometrically frustrated lattice~\cite{chern_novel_2015,grohol_spin_2005,martin_itinerant_2008,hayami_topological_2015}.
In these frustrated systems, the ground state has a large degeneracy, leaving them strongly susceptible to 
even small perturbations like longer-range exchange interactions mediated by conduction electrons coupled to 
the localized moments. In some cases, the resulting effect of the spin-charge coupling leads to 
unconventional magnetic phases~\cite{Taguchi2573,PhysRevLett.107.186403,PhysRevB.91.155132,PhysRevB.93.155115,ishizuka_loop_2013}.

Among these phases, some of the most interesting are those with noncoplanar spin orderings, 
with non-zero scalar spin chirality~\cite{PhysRevLett.105.216405,PhysRevB.89.085124,PhysRevB.93.024401}. 
The chiral nature of these states break both the parity and time-reversal symmetries. 
When an electron moves through a background of noncoplanar spin texture,  
it picks up a Berry phase, which gives rise to many interesting transport phenomena 
such as the geometric or topological Hall effect (THE) and unconventional magnetoresistive
behavior~\cite{PhysRev.95.1154,PhysRevLett.83.3737,xiao_berry_2010,Taguchi2573}. 
In THE, a transverse Hall current is observed even in the absence of any external applied magnetic field -- 
driven solely by the cumulative Berry phase acquired by the electrons. 
The acquired Berry phase is equivalent to the coupling of electron orbital moment to a fictitious magnetic field.

THE has been observed in the ferromagnetic pyrocholre compounds 
\ce{Pr2Ir2O7} and \ce{Nd2Mo2O7}~\cite{JPSJ.69.3777,PhysRevLett.98.057203,PhyRevLett.111.036602,Taguchi2573}. 
The chiral spin ordering has been studied theoretically in the context of Kondo lattice model on frustrated lattices 
such as triangular~\cite{PhysRevLett.105.266405,PhysRevB.88.235101}, 
kagom{\' e}~\cite{PhysRevB.93.024401,PhysRevB.62.R6065,barros_exotic_2014,chern_quantum_2014,PhysRevB.99.035163}, 
pyrocholre~\cite{PhysRevLett.105.226403}, face-centered cubic lattice~\cite{PhysRevLett.87.116801}, 
and checkerboard lattice~\cite{venderbos_switchable_2012}. 
Our plan is to extend this study to the geometrically frustrated SS lattice, 
which is a prototypical model of several materials like the rare-earth tetraborides
~\cite{PhysRevB.93.174408,PhysRevLett.101.177201,PhysRevB.92.214433,PhysRevB.82.214404,IGA2007e443,JPSJ.78.024707}. 
These materials have rare-earth elements with large magnetic moments that can be treated as classical spins. 
This, in turn, renders the theoretical modeling of such systems more tractable. 
For classical spins, the Kondo lattice and double exchange models can be mapped on to one another, 
as the eigenstates corresponding to opposite signs of the Kondo coupling are related by a global gauge transformation.

The SS lattice has several competing interactions in play owing to its unique lattice symmetry.
The competition between the axial and diagonal exchange interactions usually results in 
a collinear or coplanar ordered phase~\cite{PhysRevB.87.144419,PhysRevB.79.144401,SRIRAMSHASTRY19811069}. 
However, a rich variety of phases, including noncoplanar phases,
are expected when the symmetry allowed DM interaction is taken into account. 
Further, the use of an external Zeeman field enhances the possibility of having noncoplanar phases significantly.
Previously, we have shown that the Kondo lattice model on the SS lattice exhibits noncoplanar and noncollinear ground states 
over a wide ranges of parameters~\cite{PhysRevB.96.224401,PhysRevB.96.224402,Shahzad_2017}. 
In this work, we aim to thoroughly study the effect of all the competing interactions in stabilizing
the noncoplanar phases and investigate the transport properties of itinerant electrons on this lattice.
The ability to realize multiple noncollinear and noncoplanar magnetic orderings by tuning 
different interactions for realistic values of model parameters make the SS lattice an ideal case for studying THE.

In this work, we demonstrate that multiple noncollinear and noncoplanar magnetic ground state phases 
are stabilized in the SS lattice for different ranges of Hamiltonian parameters. 
The behavior of itinerant electrons is significantly modified by the coupling to the underlying spin textures. 
In particular, for noncoplanar magnetic orderings, this is manifested in the form of finite THE.

This paper is organized as follows. 
Following the introduction in section \ref{sec:intro}, 
we discuss the models used in this study in section \ref{sec:model}.
In section \ref{sec:methods} we describe the method and the observables
we calculate to characterize the magnetic and the transport properties. 
We present the results of our work in section \ref{sec:results}, 
followed by the summary in section \ref{sec:summary}.

\section{Model}\label{sec:model}
						
\begin{figure}[b]
\centering
\includegraphics[width=8.5cm,height=5.5cm]{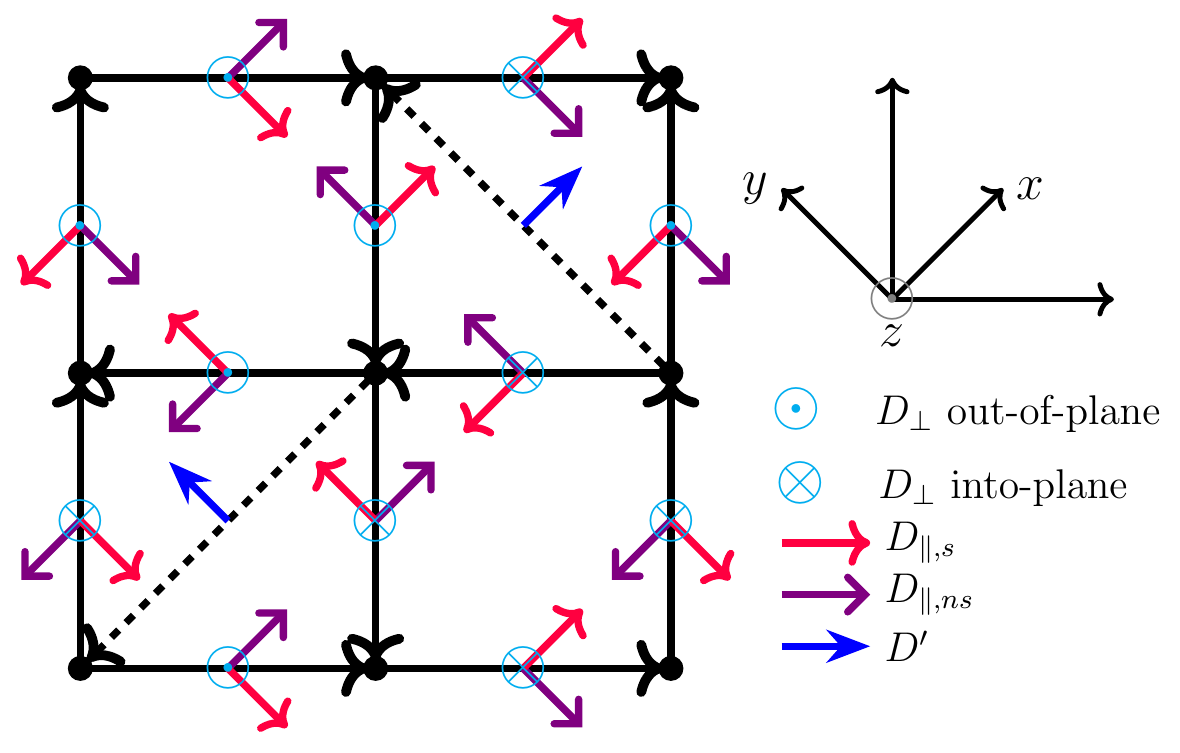}
\caption{\label{fig:ssl}
(Color online) The geometry of the SS lattice used in our study. 
Black lines represent the axial bonds while dotted black lines represent the diagonal bonds on alternate plaquette. 
The direction of arrow on these bonds indicates the order of cross product, $\mathbf{S}_i \times \mathbf{S}_j$ 
in DM term for these bonds. The in-plane component of DM vector on axial bonds 
is divided into staggered, $D_{\parallel,s}$ and non-staggered, $D_{\parallel,ns}$ components, 
and represented by red and purple arrows, respectively. 
The in-plane component of DM vector on diagonal bonds, $D'$ is indicated by blue arrows. 
The perpendicular component of DM vector, $D_{\perp}$ on axial bonds has out-of-plane and into-plane components. 
The directions of all these DM vectors are obtained using Moriya rules and crystal structure of SS lattice.
}
\end{figure}

We study the Hamiltonian,
	
\begin{equation}
\label{eq:class-ham}
\mathcal{\hat{H}}^c= \sum_{\avg{ij}}J_{ij}\mathbf{S}_i\cdot\mathbf{S}_j+\sum_{\avg{ij}}\mathbf{D}_{ij}\cdot \mathbf{S}_i\times \mathbf{S}_j-B\sum_i S_i^z
\end{equation}
     
\noindent on the SS lattice, where, $\avg{ij}$ refers to nearest neighbor axial bonds on each plaquette, 
and next nearest neighbor diagonal bonds on alternate plaquettes. 
The first term represents the Heisenberg exchange interaction, 
with $J_{ij} = J (J')$ denoting the strength of antiferromagnetic exchange on the axial (diagonal) bonds. 
The second term is the antisymmetric DM interaction with $\mathbf{D}_{ij}$ representing the DM vectors on SS bonds. 
The exact values and directions of these vectors are determined by the crystal structure, 
subject to the Moriya rules and the constraints imposed by the geometry of the lattice. 
In Fig.~\ref{fig:ssl}, the unit cell of the SS lattice together with the choice of all DM vectors on each bond is shown. 
We parameterize the DM vectors via their parallel ($D_{\parallel,s}$, $D_{\parallel,ns}$, $D'$)
and perpendicular ($D_{\perp}$ ) components.  
Further details on different components of the DM vectors are described in Fig.~\ref{fig:ssl}. 
The last term is the Zeeman coupling between localized spins and an external applied magnetic field. 

We treat the localized spins as classical vectors (true for $f$-electron systems with large magnetic moments) 
with unit length ($\vert \mathbf{S}_i\vert = 1$). 
We use the spherical polar co-ordinates, 
${\mathbf{S}_i}=(\sin\theta_i\cos\phi_i,\sin\theta_i\sin\phi_i,\cos\theta_i)$ to denote the state of the localized spin.
Henceforth, interactions on the diagonal bonds are represented with prime parameters while that on axial bonds with unprimed ones.
     
In order to study transport properties of itinerant electrons coupled to localized spin textures, we use the Kondo lattice model,
\begin{equation}
\label{eq:elec-ham}
\mathcal{\hat{H}}^{e}=-\sum_{\avg{ij},\sigma}t_{ij}(c_{i,\sigma}^\dagger c_{j,\sigma}+\mbox{H.c.})
     +J_K \sum_i \mathbf{S}_i\cdot \mathbf{s}_i
\end{equation}
\noindent where $t_{ij}$ represents the hopping matrix elements of conduction electrons on the SS lattice bonds, 
and $J_K > 0$, is the coupling strength of on-site Kondo term between classical spin, $\mathbf{S}_i$ 
and the spin of itinerant electron, $\mathbf{s}_i =  c^{\dagger}_{i\alpha} \sigma^{\alpha\beta} c_{i\beta}$. 
$J_K/t \neq 0$, lifts the spin degeneracy of conduction electron states.  
In the limit of $J_K \gg t$, the electron bands form two blocks separated 
separated by a gap $\sim J_K$ corresponding to electron spins  
aligned parallel and anti-parallel to  the localized moments,
with the spin antiparallel states occupying the lower energy bands compared to the spin parallel states.
In this limit, Hamiltonian~(\ref{eq:elec-ham}) reduces to an effective tight-binding model~\cite{RevModPhys.82.1539} for the lower energy bands, 
given by,
\begin{equation}
\label{eq:eff-ham}
{\mathcal{\hat{H}}_e}= -\sum_{\avg{i,j},\sigma}t_{ij}^{eff}(d_{i}^\dagger d_{j}+\mbox{H.c.})
\end{equation}
\noindent where,
\begin{equation}
t_{ij}^{eff}=t_{ij}e^{ia_{ij}}\cos\frac{\theta_{ij}}{2}
\end{equation}
\noindent is the effective hopping matrix elements for the spin-parallel electrons between sites $i$ and $j$.
The phase factor, related to spin chirality, is calculated as
\begin{equation}
a_{ij}=\arctan\frac{-\sin(\phi_i-\phi_j)}{\cos(\phi_i-\phi_j)+\cot\frac{\theta_i}{2}\cot\frac{\theta_j}{2}}
\end{equation}
\noindent and $\theta_{ij}$ is the angle difference between the localized spins $\mathbf{S}_i$ and $\mathbf{S}_j$,
\begin{equation}
\cos\theta_{ij}=\cos\theta_{i}\cos\theta_{j}+\sin\theta_{i}\sin\theta_{j}\cos(\phi_{i}-\phi_j).
\end{equation}
		
\section{Method and observables}\label{sec:methods}
				
To investigate the model in \eqref{eq:class-ham}, 
we use a Markov chain Monte Carlo (MC) to perform an importance sampling of the spin configurations, 
based on the Metropolis algorithm. 
The simulations are performed on lattices of dimension $L\times L$ with $L=16-48$ 
over a wide range of Hamiltonian parameters. 
We use simulated annealing procedure to prevent the freezing of the localized moments which may happen at low temperatures. 
In this approach, we start the simulations with a random spin configuration at a high temperature ($T\approx J$), 
and equilibrate the system at this temperature. Next, we decrease the temperature by $\Delta T$ 
and use the equilibrated spin configuration from previous $T$ as an initial configuration for equilibration at the new temperature. 
We repeat this process until we reach $T=0.001 J$, 
where measurements are made to calculate the thermal averages of the physical observables. 
$100\:000$ MC steps are used at each $T$ value as equilibration steps and 
further $50\:000$ MC steps are used to perform the measurements of the observables.

In order to identify the magnetic order of localized spins, 
we calculate the static spin structure factor given by the Fourier transform of spin-spin correlation function,		
\begin{equation}
\label{equ:str-fact}
S(\mathbf{Q})=\frac{1}{N^2}\sum_{i,j}\avg {\mathbf{S}_i \cdot \mathbf{S}_j} \exp [i\mathbf{Q} \cdot \mathbf{r}_{ij}],
\end{equation}
\noindent where $\mathbf{r}_{ij} = \mathbf{r}_{j} - \mathbf{r}_{i}$ denotes the position vector 
from the $\mathit{i}-$th to $\mathit{j}-$th site, 
and $\avg{\cdot}$ represents the average over different MC configurations. 
Further, to distinguish between the coplanar and noncoplanar magnetic order, 
we calculate scalar spin chirality, as a measure of noncoplanarity of spin textures. 
On a triangular plaquette, the scalar spin chirality is defined as,
\begin{equation}
\label{equ:chirality}
\chi_\bigtriangleup=\mathbf{S}_i \cdot (\mathbf{S}_{j} \times \mathbf{S}_{k}).
\end{equation} 
\noindent The total chirality $\chi$ is calculated by 
$\chi=\frac{1}{N_u}\sum_{\bigtriangleup} \chi_\bigtriangleup$, where $N_u$ is the number of SS unit cells. 
For collinear order (ferromagnetic and antiferromagnetic) and coplanar order (such as flux states), $\chi = 0$;
whereas noncoplanar magnetic ordered phases such as canted-flux, all-in/all-out and 3in-1out/3out-1in phases 
are characterized by nonzero values of $\chi$.

We use the Kubo formula to calculate the electronic transport on the magnetic ordered backgrounds
on the SS lattice. In the limit $J_K/t \rightarrow \infty$, we can use the translational invariance of the effective tight binding Hamiltonian \eqref{eq:eff-ham} to calculate the
momentum space Hamiltonian and obtain
the energy spectrum of itinerant electrons moving on a background ordered phase. 
We calculate the transverse conductivity in $\bm{k}$-space as,

\begin{equation}
\label{equ:trans-cond}
\sigma_{xy}=\frac{ie^2\hbar}{N}\sum_{m,n\neq m, \bm{k}}\left[f(\mathcal{E}_{m\bm{k}})-f(\mathcal{E}_{n\bm{k}})\right]\frac{\bra{m\bm{k}}v_x\ket{n\bm{k}}\bra{n\bm{k}}v_y\ket{m\bm{k}}}{(\mathcal{E}_{m\bm{k}}-\mathcal{E}_{n\bm{k}})^2+\eta^2}
\end{equation}
\noindent where $m$ and $n$ represent the band indices and 
$f(\mathcal{E}_{m\bm(n){k}})$ is the Fermi-Dirac distribution function for energy $\mathcal{E}_{m\bm(n){k}}$. 
$\ket{m\bm{k}}$ and $\ket{n\bm{k}}$ are eigenstates in $\bm{k}$-space corresponding to energies 
$\mathcal{E}_{m\bm{k}}$ and $\mathcal{E}_{n\bm{k}}$, respectively.
$N=L_x \times L_y$ represents the size of the sample and $\eta$ is the scattering rate. 
$v_x$ and $v_y$ are the velocity operators in $k_x$ and $k_y$ directions and can be expressed as,
\begin{equation}
v_\mu=\frac{\partial \mathcal{\hat{H}}_e}{\partial k_{\hat{\mu}}}, \quad \mu=x,y
\end{equation}

For finite values of the Kondo coupling, we diagonalize Hamiltonian \eqref{eq:elec-ham} for finite system sizes 
to obtain the energy spectrum of itinerant electrons moving on a background ordered phase. 
In this case, we calculate the transverse conductivity in $\bm{r}$-space as,

\begin{equation}
\label{equ:trans-cond2}
\sigma_{xy}=\frac{ie^2\hbar}{N}\sum_{m,n\neq m} \left[f(\mathcal{E}_m)-f(\mathcal{E}_n)\right]
\frac{\bra{m}v_x\ket{n}\bra{n}v_y\ket{m}}{(\mathcal{E}_m-\mathcal{E}_n)^2+\eta^2}
\end{equation}
\noindent where $\ket{m}$, $\ket{n}$ are single-particle eigenstates 
corresponding to energies $\mathcal{E}_m$ and $\mathcal{E}_n$, and
$v_x$ and $v_y$ are calculated as,
\begin{equation}
v_\mu=\frac{i}{\hbar}\sum_{j,\sigma}(t_{j,j+\hat{\mu}} c_{j,\sigma}^\dagger c_{j+\hat{\mu},\sigma}- \mbox{H.c.}), \quad \mu=x,y
\end{equation}

In addition to the transverse charge conductivity, we also calculate the 
transverse spin conductivity, given by an analogous Kubo formula 
that involves the spin current as,

\begin{equation}
\sigma_{xy}^{S}= \frac{ie}{4\pi N} \sum_{m,n\neq m} \left[f(\mathcal{E}_m)-f(\mathcal{E}_n)\right]
\frac{\bra{m}J_x\ket{n}\bra{n}v_y\ket{m}}{(\mathcal{E}_m-\mathcal{E}_n)^2+\eta^2}
\end{equation}
\noindent where 
$J_x = \frac{1}{2} \{v_x, diag({\bf S}_1\cdot {\bf \sigma}, ... , {\bf S}_N\cdot {\bf \sigma}) \}$ 
is the spin current operator. As we shall discuss later, 
for $J_K \sim O(t)$, 
the spin Hall conductivity exhibits characteristic features very 
different from the charge Hall conductivity.

\begin{figure}[t]
\centering
\includegraphics[width=6cm,height=9cm]{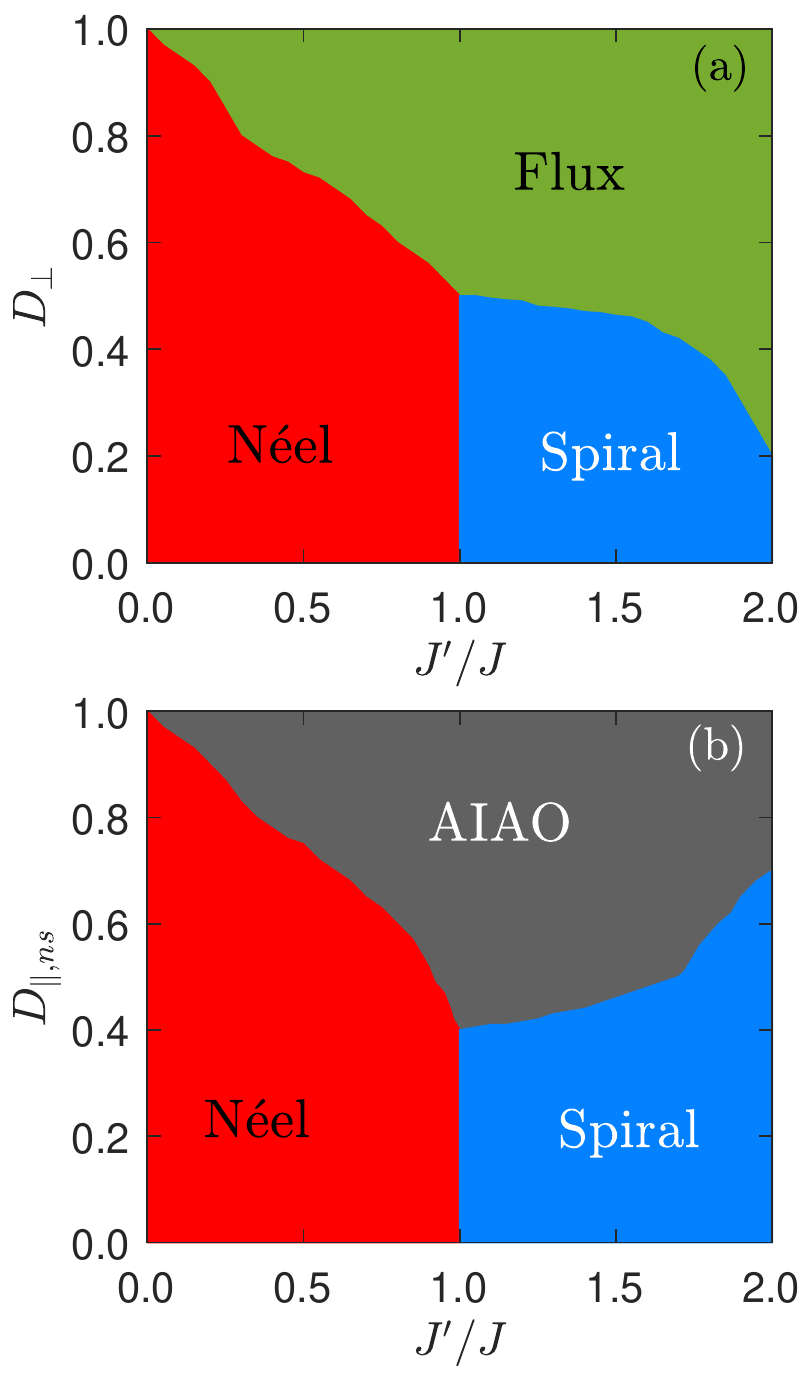}
\caption{\label{fig:phase-diags}
(Color online) Different magnetic ordered phases obtained in our simulation
with varying $J'/J$ and the DM vector components, $D_{\perp}$ (a), and $D_{||, ns}$ (b).
The other components of the DM vector, 
$D'$ or $D_{\parallel,s}$, do not stabilize any additional new phases.
}
\end{figure}

\section{Results}\label{sec:results}
\subsection{Magnetic properties}

\begin{figure*}[t]
\centering
\includegraphics[width=15.cm,height=10.0cm]{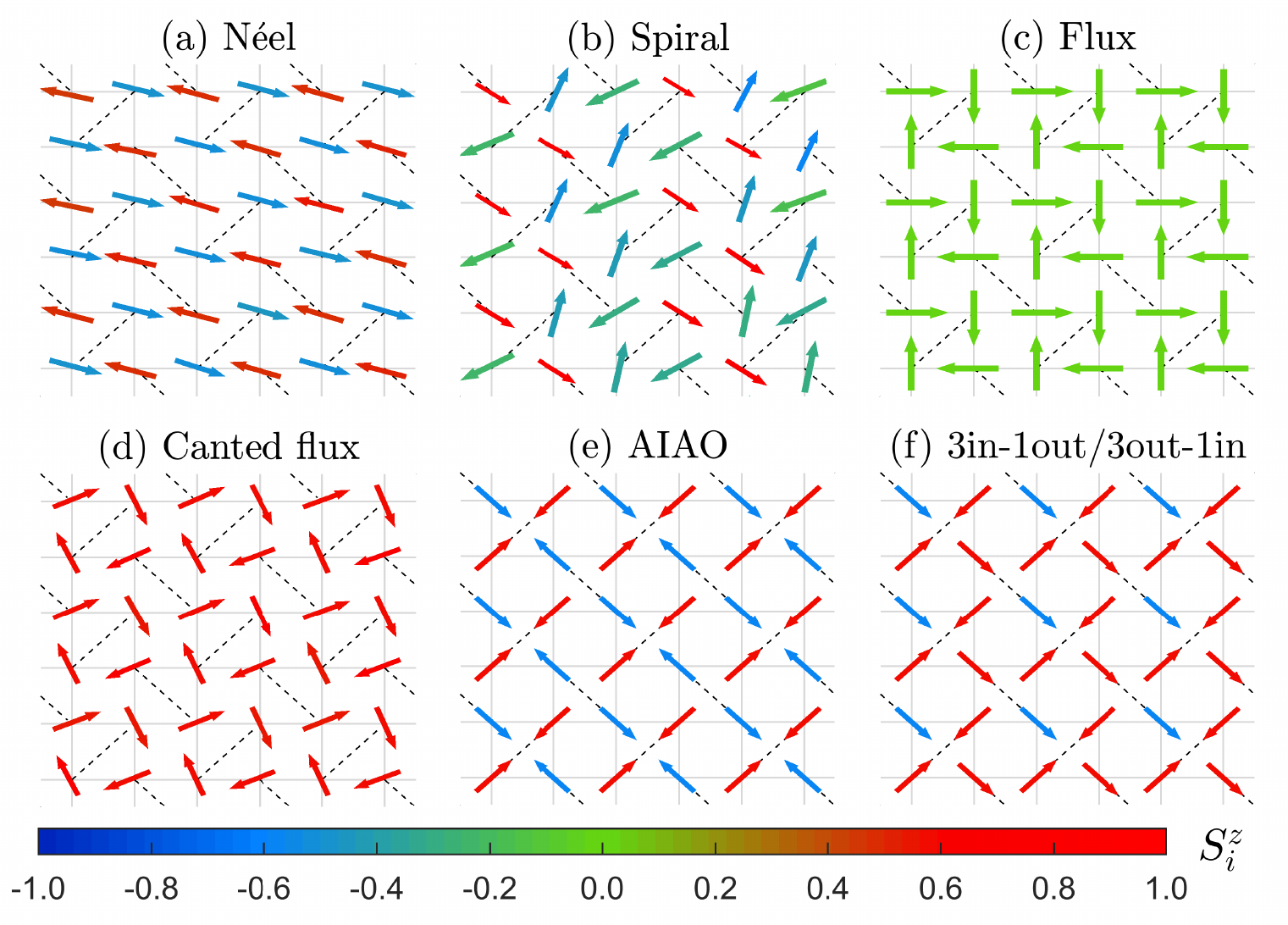}
\caption{\label{fig:configs}
(Color online) The snapshots of real-space spin configurations of localized spins $\{ {\bf S}_i \}$,
for different magnetic ordered phases on the SS lattice as seen in our MC simulations. 
The $xy$-components of spins are represented by arrows in the $xy$ plane, 
while the $z$-component is represented by the color scale. 
Hamiltonian parameters used to obtained the different phase are,
(a)~Ne\'el state at $J'/J=0.8$, (b)~spiral state at $J'/J=2.0$, (c)~flux state at $J'/J=0.8$ and $D_{\perp}/J=0.8$,
(d)~canted flux state at $J'/J=0.8$, $D_{\perp}/J=0.8$ and $B/J=5.2$, (e)~AIAO state at $J'/J=0.8$, $D_{\perp}/J=0.8$, and $D_{\parallel,ns}/J=0.7$, and (f)~3in-1out/3out-1in state at $J'/J=0.8$, $D_{\perp}/J=0.8$, $D_{\parallel,ns}/J=0.2$ 
and $B/J=4.0$.
}
\end{figure*}
	
Hamiltonian~\eqref{eq:class-ham} exhibits a wide range of magnetic orderings 
in the ground state with varying parameters. 
We observe both collinear and noncollinear, coplanar and noncoplanar magnetic orderings
for different sets of Hamiltonian parameters. The occurrence of a wide variety of ordered phases provides the motivation
to study the motion of itinerant electrons on these backgrounds with the SS lattice geometry, 
thereby exploring the novel electronic properties on a SS lattice system. 
	
We begin our study by tuning the frustration parameter $J'/J$, and different components of the DM vectors  
on the SS lattice [as shown in Fig.~\ref{fig:ssl}], in a systematic manner to identify the magnetic phase diagram in the parameter space spanned by $J'/J$ and the parallel and perpendicular components of the DM
vector. 
Fig.~\ref{fig:phase-diags} summarizes the results of our simulations.
In Fig.~\ref{fig:configs}, we show the representative spin configurations of the principal ordered phases
observed in our simulation in different parameter regimes.

In the absence of DM interaction, the ground state is a N{\' e}el antiferromagnet for $J'/J \leq 1$, and
evolves to
a spiral phase for $J'/J\gtrsim 1$ [Fig. \ref{fig:phase-diags}]. For this spiral phase, the angle difference between NN spins is $\theta =\pi\pm\cos^{-1}(J/J')$ for $J'/J >1$~\cite{SRIRAMSHASTRY19811069,PhysRevB.87.144419,PhysRevB.79.144401}.
This can be understood as  a consequence of the destabilization of the antiferromagnetic N{\' e}el state 
due to increasing frustration on the SS lattice.
With the introduction of DM component, $D_{\perp}$, there is a further competition to lower the energy 
of the spin configuration by perpendicular alignment of neighboring spins favored by the DM interaction term. Our results show that the spiral phase 
and the N{\' e}el phases are replaced by a coplanar ``flux" phase with increasing $D_{\perp}/J$
[Fig.~\ref{fig:phase-diags}(a)]~\cite{PhysRevLett.81.5604,PhysRevB.62.13816,PhysRevB.96.224401,PhysRevB.96.224402}.  Upon the inclusion of $D_{\parallel,ns}$, 
a noncoplanar all-in/all-out (AIAO) phase is observed in the ground state for intermediate to strong values of $D_{\parallel,ns}/J$
(fig. \ref{fig:phase-diags}(b)).  
The other components of the DM vector, 
$D'$ or $D_{\parallel,s}$, do not stabilize any additional phases. 
The boundaries between the different phases can be obtained
from the level crossing of the ground state energy with the variation of the parameters in the Hamiltonian.

\begin{figure*}[t]
\centering
\includegraphics[width=15.0cm,height=7.8cm]{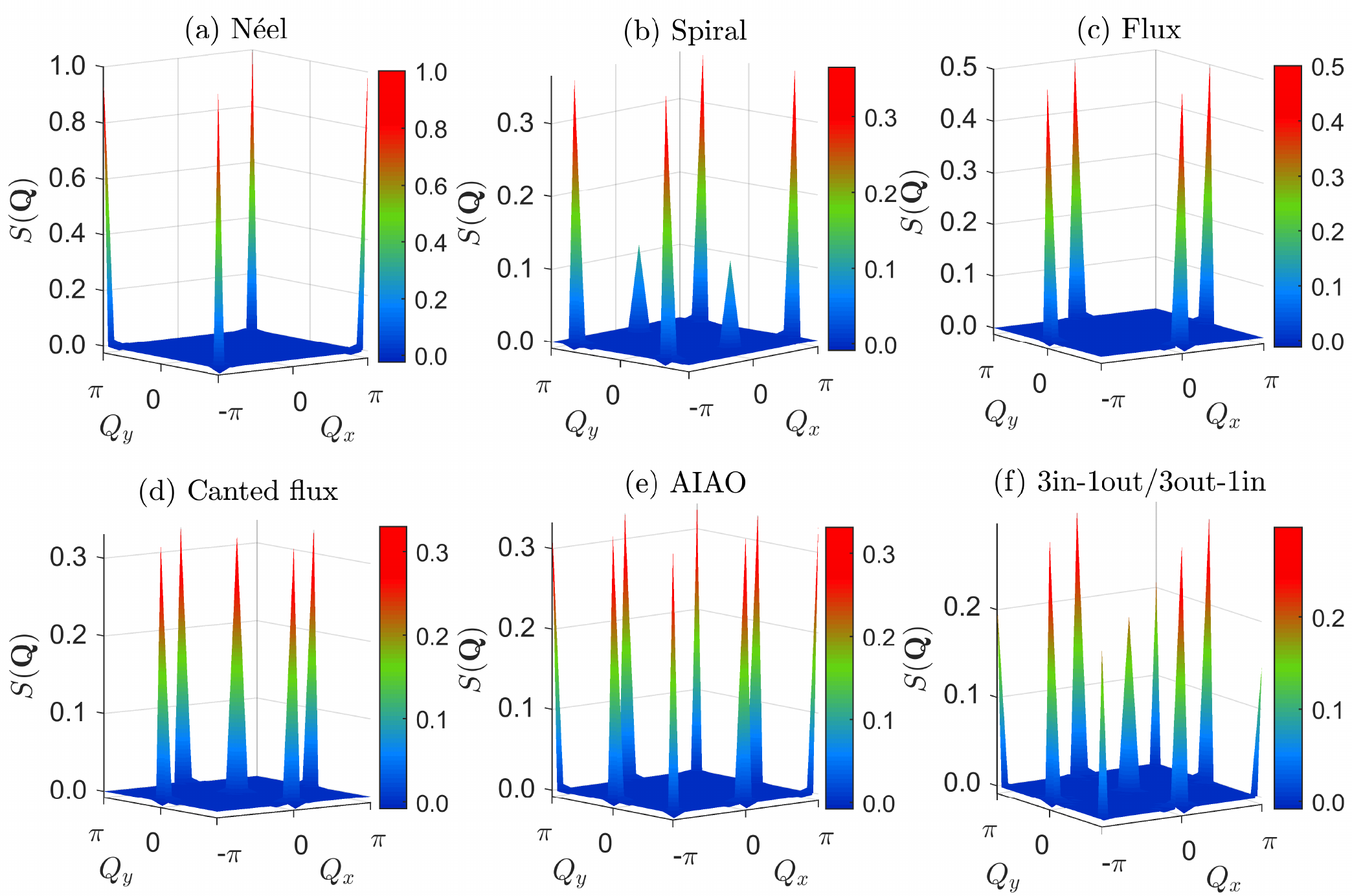}
\caption{\label{fig:sf}
(Color online) Spin structure factor, $S({\bf Q})$ showing sharp peaks for different magnetic ordered phases 
on the SS lattice as shown in Fig~\ref{fig:configs}. $S({\bf Q})$ is calculated for $Q_{x}$, $Q_{y}$ $\in$ $[-\pi,\pi]$
and the weight is represented by the color scales. We characterize the ordered phases based on their peak locations.
We observe $S({\bf Q})$ peaks at (a)~$\mathbf{Q}=(\pi,\pi)$ for the N\'eel state, 
(b)~$\mathbf{Q}=(2\pi/3,0)$ and $(2\pi/3,\pi)$ for the spiral phase, 
(c)~$\mathbf{Q}=(0,\pi)$ and $(\pi,0)$, for the flux phase, 
(d)~$\mathbf{Q}=(0,0)$, $(0,\pi)$ and $(\pi,0)$ for the canted flux phase, 
(e)~$\mathbf{Q}=(0,\pi)$, $(\pi,0)$ and $(\pi,\pi)$ for the AIAO state, and 
(f)~$\mathbf{Q}=(0,0)$, $(0,\pi)$, $(\pi,0)$ and $(\pi,\pi)$ for the 3in-1out/3out-1in phase.
}
\end{figure*}
	
Next, we explore the effect of an external magnetic field on these ordered phases on the SS lattice.
While it is tempting to map out the details of the evolution of all these candidate phases
in the presence of an external magnetic field; we restrict ourselves to the regime $J'/J =0.8$
(mainly due to the large parameter space of our model). The choice of this frustration parameter 
is motivated by experimental observation of nearly equal bond lengths in rare-earth compounds~\cite{PhysRevLett.101.177201,PhysRevB.95.174405,PhysRevB.93.174408,PhysRevB.92.214433}.
We characterize the various ordered phases and study their evolution 
in the presence of the external magnetic field by focusing on observables
such as spin structure factor and the scalar spin-chirality.
	
{\bf Structure factor :} 
A detailed understanding of the multiple magnetic states is provided by the magnetic structure factor, 
which quantifies the long range magnetic order in terms of prominent peaks in the momentum space. 
In Fig.~\ref{fig:sf}, we show the structure factor of different magnetic ordered phases observed in our simulation 
in different parameter regimes. We use the extended Brillouin zone for the spin structure factor calculation. 
We can identify different ordered phases by the location and number of peaks 
observed in the spin structure factor. 
Since, there is no spontaneous symmetry breaking in finite size systems,
we have examined the individual components of the structure factor ($\langle S_i^\mu S_j^\mu \rangle, \mu=x,y,z$), 
and the real-space spin configuration obtained from the snapshots of the MC simulation
to complement the total spin structure factor and to determine the multi-{\bf Q} ordered phases. We observe the following features,

$(i)$~In the absence of DM interaction and Zeeman field, the ground state shows 
an antiferromagnetic N\'eel ordering [see Fig.~\ref{fig:configs}(a)]. 
This can be verified from Fig.~\ref{fig:sf}(a) where the peak in spin structure factor appears at $\mathbf{Q}=(\pi,\pi)$. 
With increasing $D_\perp$, the ground state remains N\'eel antiferromagnet (AFM) 
until we reach a critical value $D_\perp^c \approx 0.62$, 
where we observe a phase transition marked by the sharp increase in the magnitude of the peak at $(0,\pi)$. 
The true nature of this ground state is revealed by the static spin structure factor shown in Fig.~\ref{fig:sf}(c), 
that exhibits two equal magnitude peaks at $\mathbf{Q}=(0,\pi)$ and $(\pi,0)$ indicating a $2{\bf Q}$ state. 
This is a noncollinear, coplanar flux state (see Fig.~\ref{fig:configs}(c)). 
Thus the system undergoes a phase transition from $1{\bf Q}$ phase (N{\'e}el state) to a $2{\bf Q}$ phase (flux state) 
with increasing $D_{\perp}$.

\begin{figure}[b]
\centering
\includegraphics[width=8.5cm,height=4.5cm]{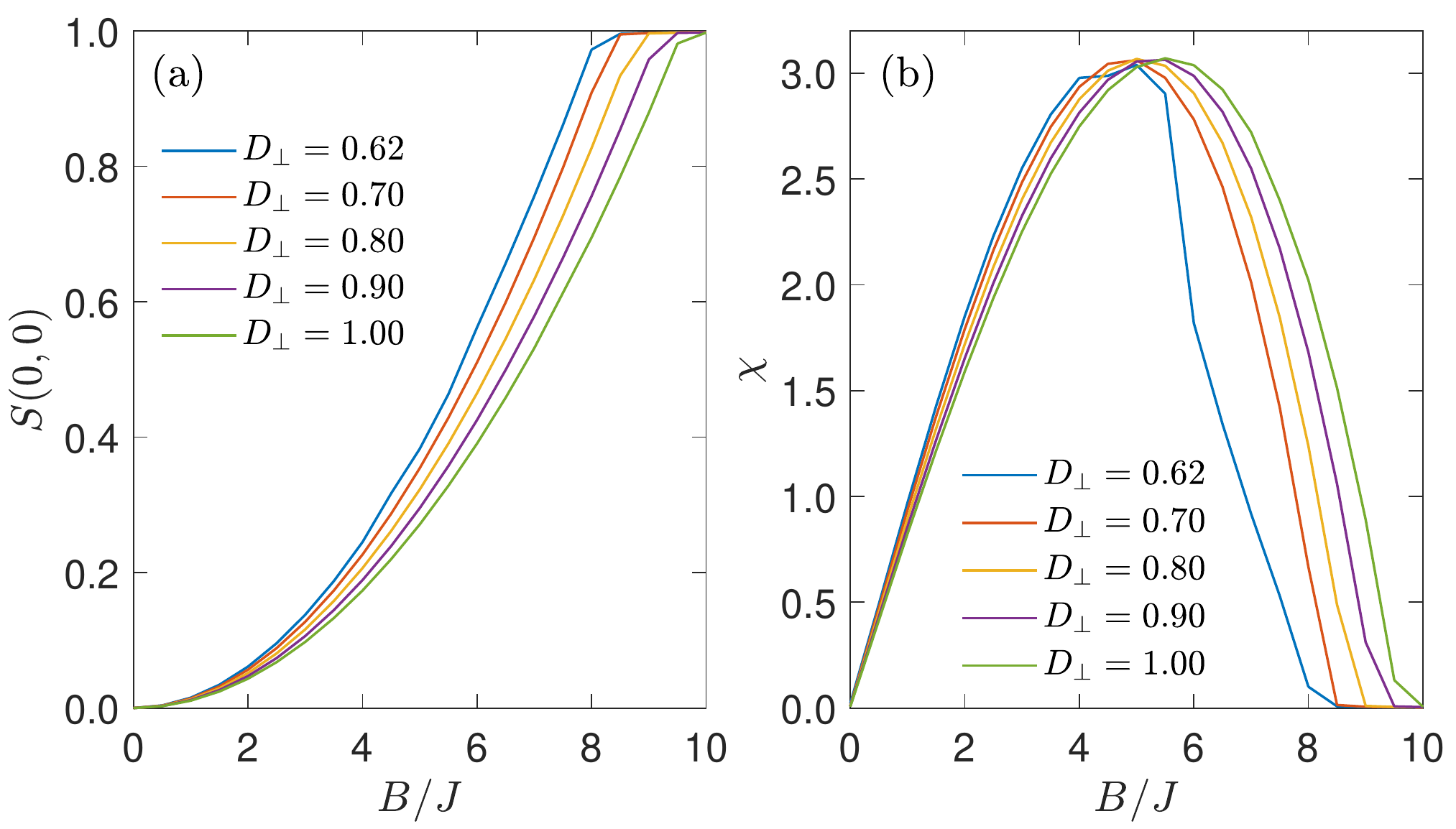}
 \caption{\label{fig:DperB}	
(Color online) (a)~Variation of the structure factor peak height at $\mathbf{Q}=(0,0)$ 
on the flux ground state
with varying external magnetic field, $B$, for fixed values of DM vector component $D_\perp$. 
$S(0,0)$ is finite for finite $B$ values, and it increases monotonically with increasing $B/J$ and 
 eventually saturating for $B/J \sim 10$.
(b)~Behavior of the scalar spin chirality, $\chi$, with varying $B$ for fixed values of $D_\perp$.
$\chi = 0$ refers to a collinear/coplanar phase where as $\chi \neq 0$ denotes a non-coplanar phase.
$\chi$ increases rapidly with increasing $B/J$, reaching a maximum for $B/J \approx 5$, and 
then reduces gradually with further increase in $B/J$.
}
 \end{figure}

$(ii)$~Next,  
the evolution of the magnetic ordering for the flux state in the presence of an external magnetic field, $B$, is investigated for an illustrative value of $D_\perp$ ($>0.62$)
\footnote{A large $D_\perp$ is chosen as it amplifies the response of the magnetic field. While this is unrealistically large compared to the intrinsic DM strength in many real magnets, recent experiments  have shown that a large DM interaction can be induced in magnetic thin films by forming interfaces with heavy metals.}.  
Introduction of a magnetic field leads to the canting of localized spins along the direction of $B$-field 
[see Fig.~\ref{fig:configs}(d)], for any non-zero B, 
which results in a $3{\bf Q}$ magnetic ordering  
exhibiting three peaks (5 peaks in the extended Brillouin Zone)  in $S({\bf Q})$
at ${\bf Q} = (\pi,0)$, $(0,\pi)$ and $(0,0)$. 
We designate this as the canted flux state. 
In Fig.~\ref{fig:DperB}(a), we show the behavior of the structure factor peak at $\mathbf{Q}=(0,0)$ 
as a function $B$ for different values of $D_\perp$. 
It can be seen that $S(\mathbf{Q}=(0,0))$ increases monotonically with increasing magnetic field strength. 
For very large $B/J$, the localized moments are aligned fully in the direction of $B$-field, 
and ground state becomes a field polarized ferromagnetic state.

\begin{figure}[t]
\centering
\includegraphics[width=8.5cm,height=4.5cm]{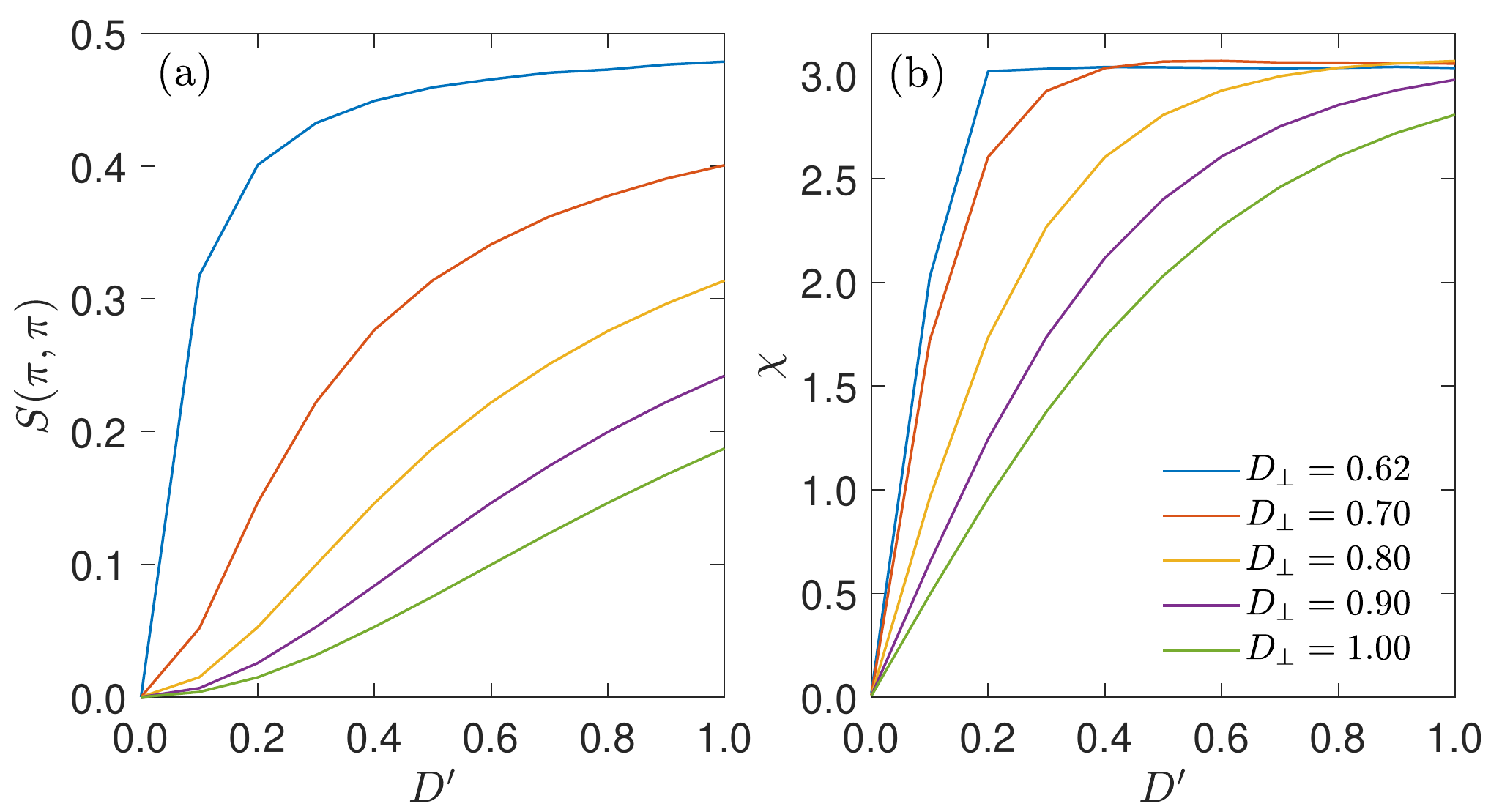}
\caption{\label{fig:Dpara}	
(Color online) Variation of $S({\bf Q})$ peak height at ${\bf Q} = (\pi,\pi)$ 
as a function of DM vector component $D'$ for fixed values of $D_\perp$.  
$S(\pi,\pi)$ increases monotonically with increasing $D'$ values.
The rise of $S(\pi,\pi)$ with $D'$ is rapid for lower values of $D_\perp$ as compared to
larger values of $D_\perp$.
(b)~Behavior of the scalar spin chirality, $\chi$, at fixed $D_\perp$ with varying $D'$.
$\chi$ increases monotonically with increasing $D'$.
However, it rises rapidly and reaches the maximum quickly for lower values of $D_\perp$.
}
\end{figure}

$(iii)$~The introduction of the parallel components of DM vector 
either on axial or on diagonal bonds also results in the canting of localized spins. 
Fig.~\ref{fig:Dpara} shows the effect of parallel component of DM vector on flux state. 
There is an additional peak in $S({\bf Q})$  at $(\pi,\pi)$ and its weight increases 
with the increase of strength of parallel component. 
Qualitatively, the effect is same for all three parallel components of DM vectors namely $D_{\parallel,s}$, 
$D_{\parallel,ns}$ and $D'$. The ground state has $3\mathbf{Q}$ magnetic ordering with peaks in $S({\bf Q})$ at $\mathbf{Q}=(0,\pi)$, $(\pi,0)$ and $(\pi,\pi)$ as shown in Fig.~\ref{fig:sf}(d).
This phase corresponds to an AIAO state, where the orientation of four neighboring spins on the SS lattice plaquettes with diagonal bonds pointing in different directions
can be mapped to the four radially inward/outward pointing vectors from the vertices 
of a regular tetrahedron (see Fig.~\ref{fig:configs}(e))
The transformation of the flux state to an AIAO state occurs for non-zero values of $D_{\parallel,ns}$ 
(or, alternatively, $D_{\parallel,s}$ or $D'$) that increases monotonically with $D_\perp$.

\begin{figure}[b]
 \centering
 \includegraphics[width=8.5cm,height=4.5cm]{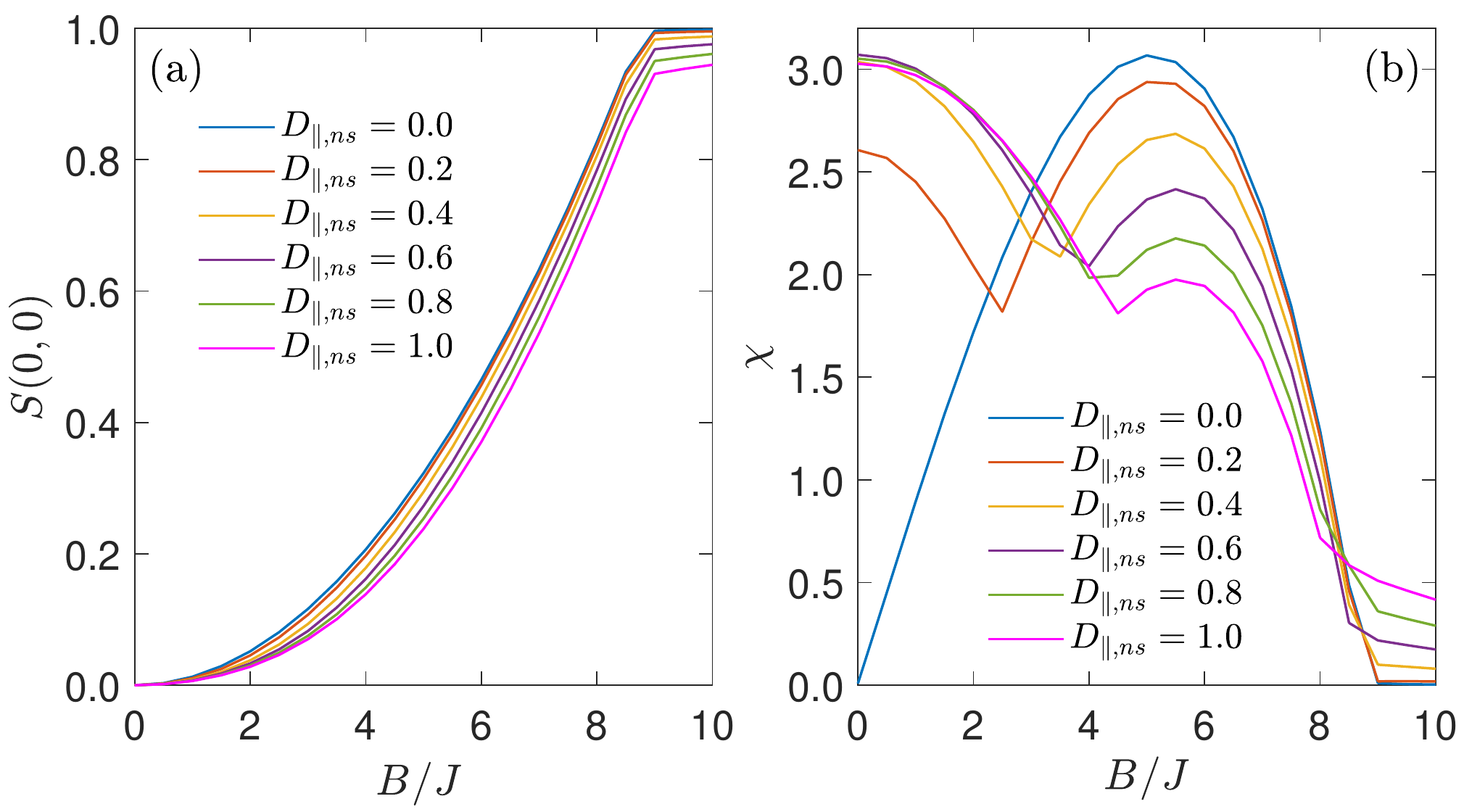}
 \caption{ \label{fig:DparaB}
 (Color online) (a)~Behavior of the structure factor peak weight at $\mathbf{Q}=(0,0)$ 
 on the AIAO ground state
 as a function of external magnetic field $B$ and the DM vector component $D_{\parallel,ns}$.
 $S(0,0)$ is finite for finite $B$ values, and it increases monotonically with increasing $B/J$ and 
 eventually saturating for $B/J \sim 10$.
 (b) Variation of scalar spin chirality with increasing $B$ at fixed values of $D_{\parallel,ns}$.
 For $D_{\parallel,ns} = 0$, with increasing $B$, $\chi$ increases rapidly, reaches a maximum and 
 then reduces gradually reaching zero for $B/J \sim 10$.
 For $D_{\parallel,ns} \neq 0$, $\chi$ is finite in the $B=0$ limit. 
 $\chi$ shows a non-monotonic behavior with increasing $B/J$ and vanishes in the limit $B/J \approx 10$.
}
 \end{figure}

$(iv)$~Next, we apply magnetic field in the presence of both in-plane and perpendicular components of DM vector. 
In the absence of magnetic field as mentioned in the previous paragraph the magnetic ordering is that of an AIAO type. 
With the introduction of an external magnetic field the localized moments reorient in the direction of the $B$-field 
and we get an additional out-of-plane canting of these moments [see Fig.~\ref{fig:configs}(f)]. 
The peak in spin structure factor at $\mathbf{Q}=(0,0)$ grows with increasing magnetic field 
as shown in Fig.~\ref{fig:DparaB}(a) on a color scale. 
The ground state now has $4\mathbf{Q}$ ordering with peaks in $S({\bf Q})$ located at 
$\mathbf{Q}=(0,0)$, $(0,\pi)$, $(\pi,0)$ and $(\pi,\pi)$ as shown in Fig.~\ref{fig:sf}(f). 
This is a  3in-1out/3out-1in state with three spins pointing in and one spin pointing out from the center of tetrahedron.
Further increase in magnetic field results in all localized spins pointing in the direction of $B$-field, 
a fully polarized ferromagnetic state.

{\bf Spin chirality :} As seen above, multiple magnetic ordered phases are stabilized in the current model 
due to the interplay of the antiferromagnetic exchange interaction, the DM interaction and the external magnetic field. 
To quantify the noncoplanarity of these spin textures we look into the scalar spin-chirality, $\chi$ (Eq. \eqref{equ:chirality}). 
Our calculation of $\chi$ gives the following results.

The N\'eel state being a collinear state has zero spin chirality.
Further, the chirality also vanishes in the spiral phase.	
The chirality of flux state is zero as it is a $2\mathbf{Q}$ noncollinear, but coplanar state. 
With increasing magnetic field on this flux state, the canting of the local moments in the direction of $B$-field 
increases continuously until the local moments are fully polarized. 
The chirality for canted flux state is non-zero as it is a noncoplanar state with $3\mathbf{Q}$ magnetic ordering. 
As shown in Fig.~\ref{fig:DparaB}(b), the chirality increases monotonically up to an intermediate value of the applied field 
and then decreases continuously to zero at saturation.  
	
Introduction of any of the parallel components of DM vectors causes the flux state 
to have an out-of-plane canting of the localized spins. For such states, 
$S({\bf Q})$ shows additional peaks at ${\bf Q} = (\pi,\pi)$. 
The weight of this peak increases with the increase of any of the in-plane component of DM vectors. 
This $3{\bf Q}$ state is an AIAO state with non-zero spin chirality.
Applying the magnetic field changes the AIAO state to a 3in-1out/3out-1in state. 
For this state, $S({\bf Q})$ shows one more peak at ${\bf Q} = (0,0)$. 
The enlarged out-of-plane component of the spins contribute to an increase in noncoplanarity of the ground state. 
The magnitude of the spin chirality increases with increasing magnetic field strength. 
The 3in-1out/3out-1in state is a $4{\bf Q}$ state with a non-zero chirality as shown in Fig.~\ref{fig:DparaB}(b).

\subsection{Electronic properties}

{\bf Band structure :}
Coupling to the local moments modifies the transport properties of 
itinerant electrons dramatically. For simplicity, we consider  
a single band of $s$-electrons interacting with the magnetic ordering via a Kondo coupling 
between the electron spin and the local moments, as given by the Hamiltonian~(\ref{eq:elec-ham}). 
The dynamics of the electrons is fast compared to that of the localized classical spins. 
Consequently, at short time scales, the electrons effectively move in a static, 
but spatially varying magnetic field. Each local moment, ${\bf S}_i$ acts as a local magnetic field 
whose action on the spin magnetic moment of the itinerant electrons ${\bf s}_i$ is described by a Kondo-like interaction
 $J_K\mathbf{S}_i\cdot \mathbf{s}_i$. 
In comparison, the Zeeman energy due the external magnetic field coupled to the spin of the electron is small 
and shall be neglected. In the following, the hopping amplitude along the axial bonds $t$ is chosen to be unity ($t=1.0$). 
For diagonal bonds the hopping matrix element is fixed at $t'/t=0.8$.

\begin{figure}[t]
\centering
\includegraphics[width=8.5cm,height=8.0cm]{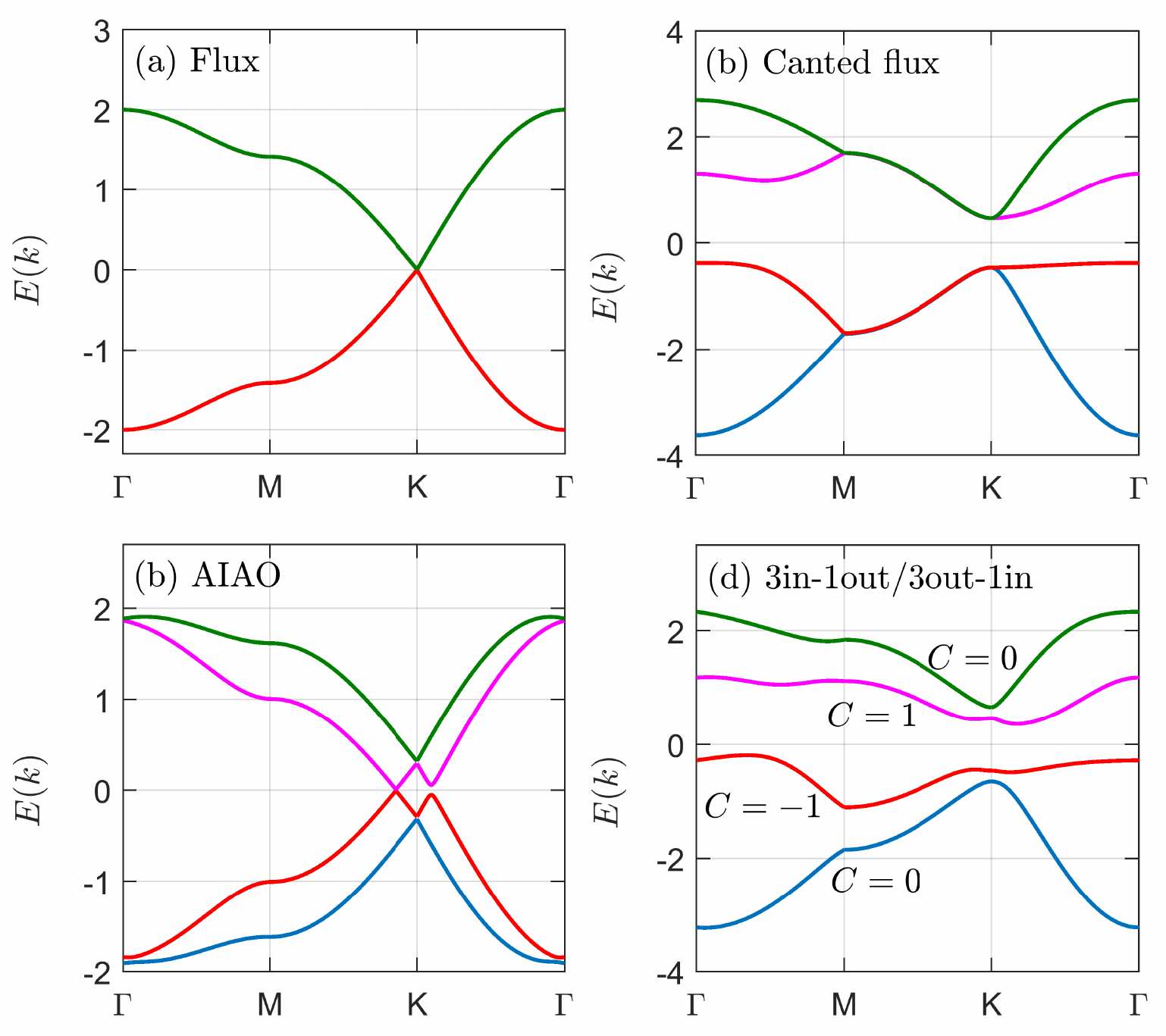}
\caption{\label{fig:band_str}
(Color online) The band structure for itinerant electrons plotted along a high symmetric path in the BZ 
for different magnetic orderings of localized spins 
(a)~flux state, (b)~canted flux state, (c)~AIAO state and (d)~3in-1out/3out-1in state. 
The ratio of hopping matrix on diagonal and axial bond is set to $t'/t=0.8$.
}
\end{figure}
	
In the absence of an external field, the electron band structure of SS lattice consists of 4 bands 
with 2-fold spin degeneracy as the SS lattice has 4-site unit cell. 
One of the bands is flat along the diagonal of the  Brillouin zone (BZ) which gives rise to strong Van Hove singularity, 
where any interaction effects are maximized. A coupling to the spin texture increases the size of the unit cell 
in accordance with the periodicity of the magnetic ordering. 
The BZ is proportionately reduced and the bands are folded into the first BZ. 
$J_K > 0$ lifts the spin degeneracy and the energy bands for electrons 
with spins anti-parallel and parallel to the local moments are shifted downwards and upwards respectively. 
For sufficiently strong Kondo-coupling, i.e., $J_K \gg t_{ij}$, 
the spin parallel and anti-parallel bands are completely separated by a gap  $2J_K$, 
and we end up with an effective tight-binding model as discussed in section~\ref{sec:model}. 
In this limit, the effective magnetic field produced by the spin texture couples directly to the charge degrees of freedom 
of the itinerant electrons, analogous to Quantum Hall systems. 
The electron energy bands are modified depending on the nature of the underlying magnetic order. 

In Fig. \ref{fig:band_str}, we show the electronic band structure along a high symmetry path in the 1st BZ,
for the four magnetic ordered phases which are stabilized in the SS lattice. 
The high-symmetry points of BZ taken in the calculations are $\Gamma=(0,0)$, $M=(\pi/2,0)$ and $K=(\pi/2,\pi/2)$. 
We observe following key features.

\begin{figure}[b]
\centering
\includegraphics[width=8.5cm,height=8.0cm]{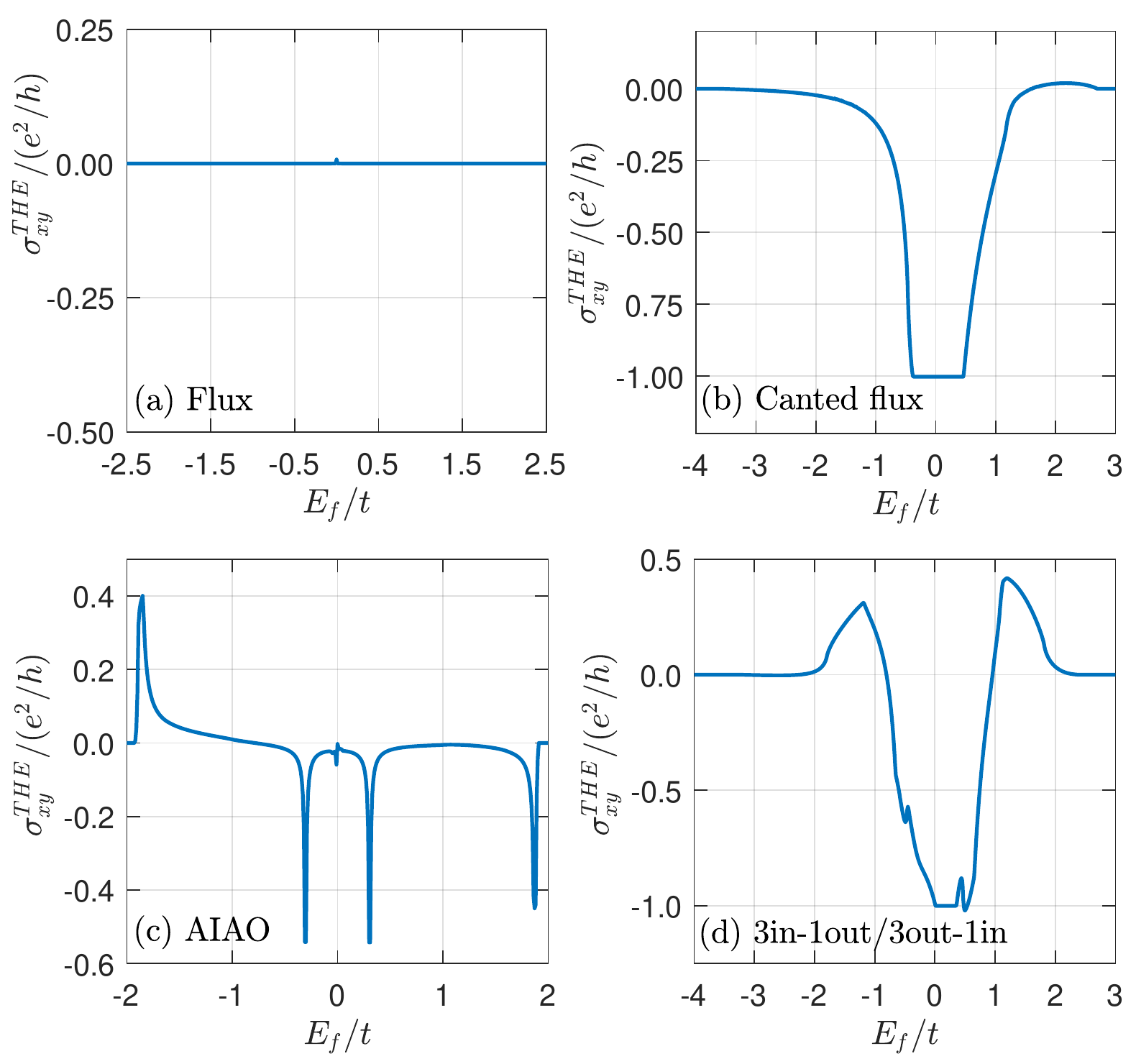}
\caption{\label{fig:hall_cond}
(Color online) Topological Hall conductivity of conduction electrons 
in the $J_K \gg t$ limit, as a function of Fermi energy, 
when they move in the background of different magnetic phases 
(a)~flux state, (b)~canted flux state, (c)~AIAO state and (d)~3in-1out/3out-1in state. 
We use $t'/t = 0.8$.
}
\end{figure}

\noindent (i) \underline{\it{Flux state}}: The magnetic unit cell of the SS lattice remains as four sites for flux type ordering 
of the localized spins. The band structure consists of eight bands and for large $J_K$ these split into four bands each 
for spin parallel and antiparallel alignment of itinerant electrons with the localized moments. 
We show the dispersion of itinerant electrons when they move on the background of flux phase in Fig.~\ref{fig:band_str}(a).
The four spin anti-parallel bands are doubly degenerate and touch each other at the $K$-point of BZ.
	
\noindent (ii) \underline{\it{Canted flux state}}: For this magnetic state, the dispersion of conduction electrons 
are plotted in Fig.~\ref{fig:band_str}(b). It consists of four bands with degeneracy of the bands is partially lifted. 
There is a gap opening between upper and lower pair of bands. 
The non coplanar canted flux state not only lifts the degeneracy, but also opens up a direct band gap at the $K$-point.
	
\noindent (iii) \underline{\it{AIAO state}}: For this state, the magnetic unit cell is also four sites.
The band structure comprises of four bands as shown in Fig.~\ref{fig:band_str}(c). 
The degeneracy of the bands is lifted and we observe an indirect gap between upper and lower pair of bands 
at the $K$-point. Interestingly, the middle two bands touch each other at a point close to the $K$-point.
	
\noindent (iv) \underline{\it{3in-1out/3out-1in state}}: The size of the magnetic unit cell remains same
as the SS lattice for this magnetic ordering. The degeneracy of all four bands is fully lifted 
and we observe direct as well as indirect band gaps between the bands as shown in Fig.~\ref{fig:band_str}(d).
We calculate the Chern number of the bands in this state
and found that two bands have non-zero Chern numbers [see Fig.~\ref{fig:band_str}(d).

\begin{figure}[t]
\centering
\includegraphics[width=8.5cm,height=7.0cm]{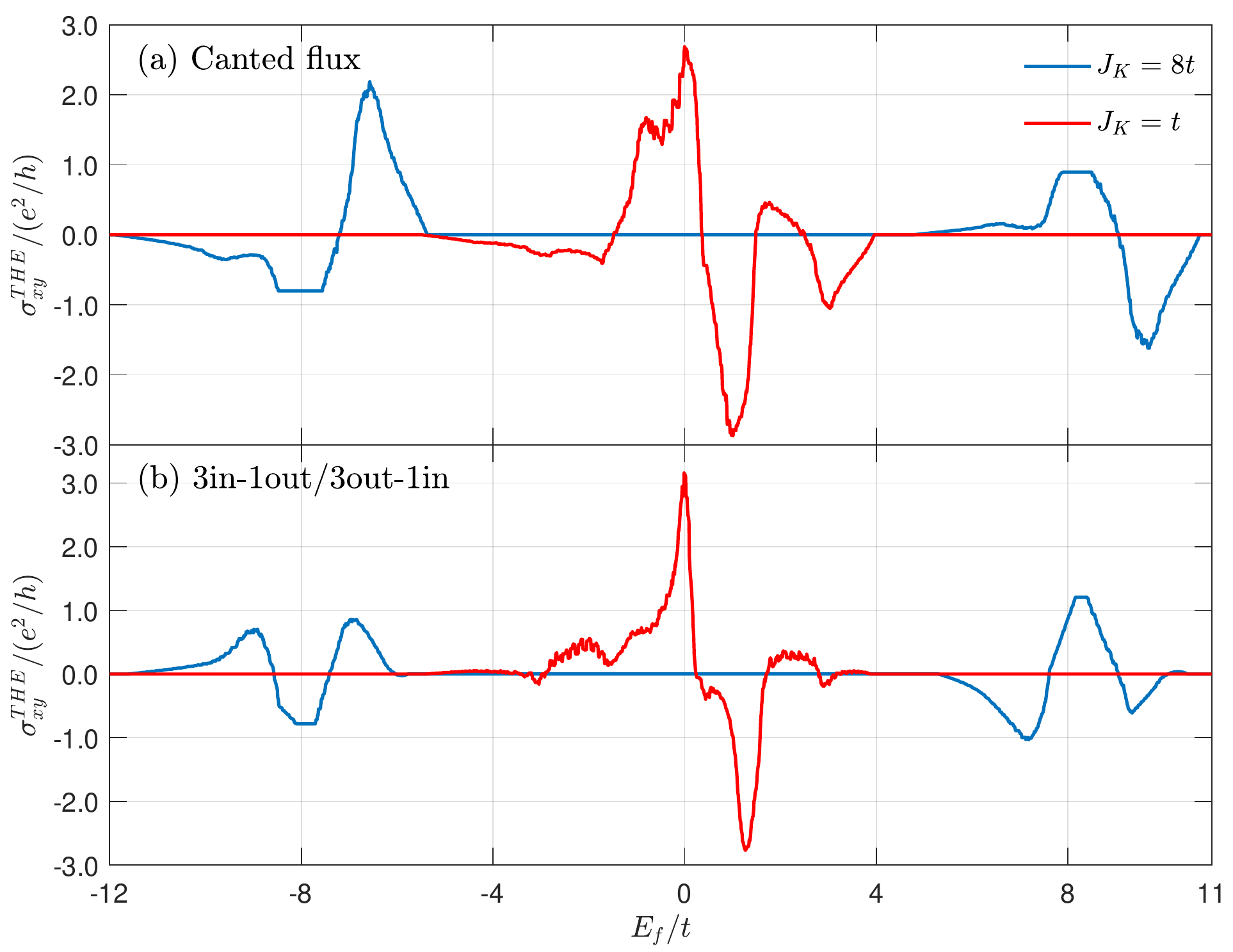}
\caption{ \label{fig:cond_jk}
(Color online) Behavior of topological Hall conductivity as a function of the Fermi energy
for $J_K = 8t$ and $t$, with conduction electrons coupled to 
(a)~canted flux state and (b)~3in-1out/3out-1in state orderings of the localized moments on the SS lattice.
}
\end{figure}

{\bf Hall conductivity at infinite Kondo coupling:} The coupling to local moments 
modifies the transport properties of itinerant electrons significantly in metallic magnets. 
The effect is most dramatic in the transverse conductivity, 
especially when the underlying spin arrangement is noncoplanar. 
In a magnetic metal, the Hall resistivity consists of three contributions,
$$ \rho_{xy} = \rho_{xy}^\text{NHE} + \rho_{xy}^\text{AHE} + \rho_{xy}^\text{THE},$$
where NHE, AHE and THE refer to Normal, Anomalous and Topological Hall effects, respectively. 
The AHE appears in metals with a net magnetization due to spin-orbit coupling. 
On the other hand, THE arises due to the Berry phase acquired by an electron 
moving in a noncoplanar spin texture. The phenomenon is best understood 
within the framework of the effective Hamiltonian (\ref{eq:eff-ham}) in the strong coupling limit ($J_K\gg t$). 
In this limit, the Berry phase acquired by an electron moving around a closed plaquette 
results in an effective flux threading each such plaquette that acts as a fictitious magnetic field 
and gives rise to a Hall effect, whose origin is purely geometrical. 
Further, it depends on the value of the Fermi energy.
In this work, we focus only on the contribution of THE to the transverse conductivity 
for different background magnetic phases with varying Fermi energy. 
In the strong coupling limit, we use the Hamiltonian (\ref{eq:eff-ham}) and 
the momentum space Kubo formalism (equation \ref{equ:trans-cond})
to study the THE. We observe the following,

\noindent (i) \underline{\it{Flux state}}:  The Hall conductivity of electrons
moving on a of background of the flux phase with the electron spin strongly coupled to the local moment 
is plotted in Fig.~\ref{fig:hall_cond}(a) for varying the chemical potential values.
We observe the Hall conductivity remains zero throughout the entire range of chemical potential. 
As identified earlier, the flux state is a coplanar state with zero chirality. 
This explains the vanishing THE for the flux state.
	
\noindent (ii) \underline{\it{Canted flux state}}: As discussed before,
the canted flux state is a $3\mathbf{Q}$ state 
and the electronic band structure displays a direct band gap at the $K$-point for this state. 
Further, the spin chirality associated with this noncoplanar state is non-zero, and that contributes to THE. 
The transverse conductivity as a function of Fermi energy for canted flux state is shown in Fig.~\ref{fig:hall_cond}(b).  
We observe a plateau in the Hall conductivity as the chemical potential falls in the band gap.
The Hall conductivity has the quantized value $-1$ (in unit of $e^2/h$).  
	
\noindent (iii) \underline{\it{AIAO state}}: The behavior of Hall conductivity with changing chemical potential 
for this phase is shown in Fig.~\ref{fig:hall_cond}(c).  
There is a non-zero value of Hall conductivity for small range of Fermi energy which is attributed 
to non-zero value of chirality for this state. The value of the conductivity is not integer 
as there is no direct band gap between the energy bands.

\begin{figure}[b]
\centering
\includegraphics[width=8.5cm,height=7.0cm]{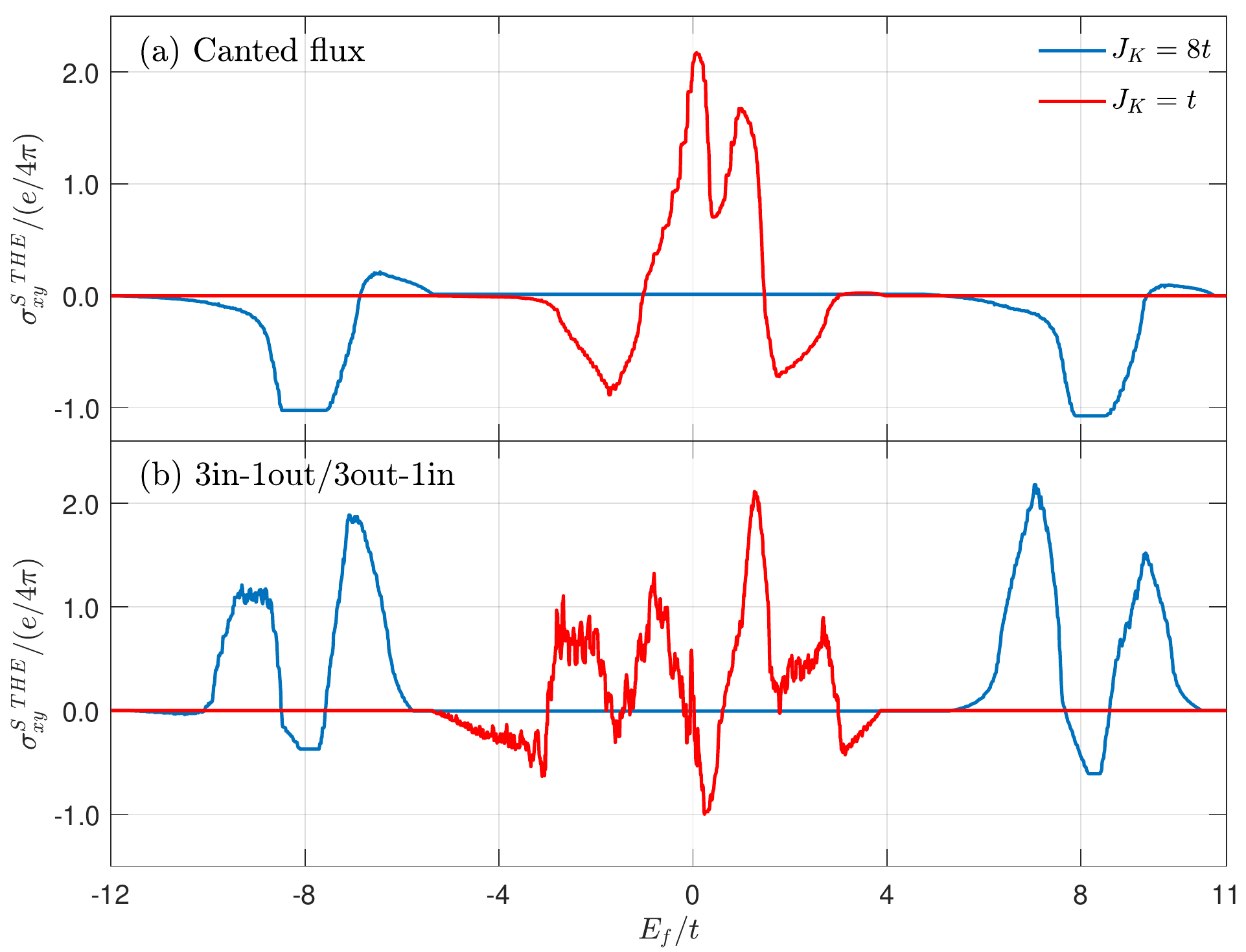}
\caption{ \label{fig:spin_cond_jk}
(Color online) Behavior of topological spin Hall conductivity as a function of the Fermi energy
for different  Kondo coupling values of conduction electrons coupled to 
(a)~canted flux state and (b)~3in-1out/3out-1in state orderings of the localized moments.
}
\end{figure}

\noindent (iv) \underline{\it{3in-1out/3out-1in state}}: The most interesting outcome of our work is observed for the
this magnetic state. This magnetic phase is noncoplanar with non-zero value of chirality. 
 We also observe that degeneracy of all the bands is fully lifted and there are direct and indirect gaps between the bands. 
 The Hall conductivity for this phase is shown in Fig.~\ref{fig:hall_cond}(d) and it remains non-zero for a 
 large window of Fermi-energy lying between $-2t$ and $2t$. Again the noncoplanarity of this phase 
 manifests itself through non-zero value of THE. When the gap between the energy bands is direct, 
 the quantized Hall conductivity remains $-1$ (in unit of $e^2/h$) for the width of band gap. 
The quantized value of $\sigma_{xy}$ is related to the sum of Chern numbers of the lowest two bands
 in the energy spectrum (see fig.\ref{fig:band_str}(d)).
 This is a signature of integer THE similar to integer quantum Hall effect observed in quantum Hall systems.

{\bf Hall conductivity at finite Kondo coupling:}
Having studied the behavior of the topological Hall effect in the ground state 
in the $J_K \gg t$ limit, we next attempt to find it at intermediate values of the Kondo coupling, $J_K \sim O(t)$.
We use the Hamiltonian (\ref{eq:elec-ham}) and the real-space Kubo formalism (equation \ref{equ:trans-cond2})
to perform our transport calculations in this regime.
Unlike the $J_K \gg t$ limit, where the contribution to the Hall conductivity is
due to the electronic states of either the spin anti-parallel or the spin parallel to the local moments, 
for $J_K \sim O(t)$, the contribution is due to the electronic states of
both spin parallel and anti-parallel to the local moments. 
Further, the contribution to topological Hall effect strongly depends on 
the value of Fermi-energy.  
 
We calculate the Hall conductivity for a canted flux state and a 3in-1out/3out-1in state,
for finite Kondo coupling values by varying the Fermi energy. 
The results are shown in Fig.~\ref{fig:cond_jk}. 
We observe that for both the spin backgrounds, 
$(i)$~the Hall conductivity due to the electrons aligned anti-parallel and parallel show 
similar contribution, but with opposite signs for $J_K = 8t$. 
This can be understood by the fact that the opposite electron spin alignment 
with respect to the local magnetic ordered phases, gives rise to emergent magnetic fields of opposite signs. 
As a manifestation of this effect, in a semi-classical picture, electrons of opposite spins 
deflect in opposite transverse directions due to the emergent magnetic fields. 
$\sigma_{xy}$ changes sign as the Fermi energy crosses a van Hove singularity. 
It exhibits a quantized value when the Fermi energy lies within the band gap
for both the ordered phases.
$(ii)$~For $J_K = t$, the Hall conductivity not only shows new features as compared to the $J_K \gg t$ limit,
but also has a large contribution even at $E_f = 0$. This new contribution to the Hall conductivity 
is due to the overlap of the electronic states aligned parallel and anti-parallel to the local spin background.   
Here, as in the previous case, $\sigma_{xy}$ changes sign as the Fermi energy crosses a van Hove singularity. 
However, $\sigma_{xy}$ does not show any quantized values over the whole range of Fermi energy 
indicating the absence of any clear band gap in the electronic states. 

Next we discuss the behavior of the topological spin Hall conductivity ($\sigma_{xy}^S$).  
While both $\sigma_{xy}$ and $\sigma_{xy}^S$ are interlinked, 
they also exhibit some distinct features which makes this study interesting.
Fig.~\ref{fig:spin_cond_jk} shows the variation of the spin Hall conductivity 
with changing chemical potential for different values of $J_K$ 
on the canted flux and the 3in-1out/3out-1in phases respectively.
It can be seen that $\sigma_{xy}^S$ is symmetric for positive and negative values 
of the chemical potential. However, this is not the case for $\sigma_{xy}$ [see fig.\ref{fig:cond_jk}]. 
In the strong coupling limit ($J_K \gg t$), the energy bands for local spin-aligned and 
anti-aligned electrons are separated by a wide band gap. 
As a result, $\sigma_{xy}$ and $\sigma_{xy}^S$ follow one another closely.
In this limit, only one species of electrons contribute to the Hall conductivities. 
Both $\sigma_{xy}$ and $\sigma_{xy}^S$, exhibit sharp jumps and change signs 
as the Fermi energy is tuned across the van Hove singularities.
For $J_K \sim t$, the local spin polarization is incomplete.
This leads to different fractions of spin parallel and spin anti-parallel states 
with strong overlap in energy of these states. 
The electron spin states hybridize and the spins of itinerant electrons 
are not simply aligned or anti-aligned to the local moments. 
The energy eigenstates have contributions from both electronic spin states.
As a consequence, the $\sigma_{xy}$ and $\sigma_{xy}^S$ are decoupled from each other.
For $J_K = t$, there exists ranges of Fermi energy for which $\sigma_{xy}^S > \sigma_{xy}$
This is suggestive of the fact that electrons with opposite spins are deflected in opposite directions, 
which leads to an increase in the spin Hall conductivity, and a decrease in the charge Hall conductivity 
as compared to the case of the zero-field non-overlapping band scenario.

\section{Summary}\label{sec:summary}				
We have identified multiple noncollinear and noncoplnar magnetic phases stabilized on the SS lattice 
in the presence of competing antiferromagnetic exchange couplings, DM interaction and an external magnetic field. 
We discuss the role of in-plane, and out-of-plane components of the DM vectors, and external magnetic field 
in the stabilization of these exotic ground states of localized moments. 
Having identified the unconventional magnetic orderings, we discuss the novel electronic properties due to the coupling of itinerant electrons to these complex spin textures, focusing on the topological Hall effect. 
Our study of the topological Hall effect on the SS lattice for strong and intermediate Kondo-couplings
between localized spins and itinerant electrons shows distinct contributions to Hall conductivities.
Our results predict occurrence of THE on the SS lattice and may be seen in experiments on rare-earth tetraborides.

\begin{acknowledgments}
We acknowledge the use of the HPCC cluster at NTU, Singapore and 
the NSCC ASPIRE1 cluster in Singapore for our numerical simulations. 
The work is partially supported by Grant No. MOE2014-T2-2-112 of the Ministry of Education, Singapore.
\end{acknowledgments}
			
\bibliographystyle{apsrev4-1}
\bibliography{klm_project}
			
\end{document}